\documentclass[12pt,preprint]{aastex}

\newcommand{\psr}{PSR~J1909$-$3744}

\shorttitle{The Mass of a Millisecond Pulsar}
\shortauthors{Jacoby et al.}

\begin{document}

\title{The Mass of a Millisecond Pulsar}

\author{B. A. Jacoby\altaffilmark{1,2}, A. Hotan\altaffilmark{3,4}, M. Bailes\altaffilmark{3}, S. Ord\altaffilmark{3,5}, and S. R. Kulkarni\altaffilmark{1}}

\altaffiltext{1}{Department of Astronomy, California Institute of
Technology, MS 105-24, Pasadena, CA 91125; srk@astro.caltech.edu.}

\altaffiltext{2}{present address: Naval Research Laboratory, Code 7213,
4555 Overlook Avenue, SW, Washington, DC 20375; bryan.jacoby@nrl.navy.mil.}

\altaffiltext{3}{Centre for Astrophysics and Supercomputing, Swinburne
University of Technology, P.O. Box 218, Hawthorn, VIC 31122,
Australia; ahotan@astro.swin.edu.au, mbailes@astro.swin.edu.au.}

\altaffiltext{4}{Australia Telescope National Facility, CSIRO,
P.O. Box 76, Epping, NSW 1710, Australia.}

\altaffiltext{5}{present address: School of Physics, University of Sydney, A28,
NSW 2006, Australia; ord@physics.usyd.edu.au.}

\begin{abstract}
We report on nearly two years of timing observations of the low-mass
binary millisecond pulsar, \psr~with the Caltech-Parkes-Swinburne
Recorder~II (CPSR2), a new instrument that gives unprecedented timing
precision. Daily observations give a weighted rms residual of 74\,ns,
indicating an extremely low level of systematic error.  We have
greatly improved upon the previous parallax and proper motion
measurements of \psr, yielding a distance of
$(1.14^{+0.04}_{-0.03})$\,kpc and transverse velocity of
$(200^{+7}_{-6})$\,km\,s$^{-1}$. The system's orbital eccentricity is
just 1.35(12)$\times10^{-7}$, the smallest yet recorded.  Since their
discovery, the masses of the rapidly rotating millisecond pulsars have
remained a mystery, with the recycling hypothesis arguing for heavy
objects, and the accretion-induced collapse of a white dwarf more
consistent with neutron stars less than the Chandrashkar
limit. Fortuitously, \psr~is an edge-on system, and our data have
allowed the measurement of the range and shape of the Shapiro delay to
high accuracy, giving the first precise determination of a millisecond
pulsar mass to date, $m_p = (1.438 \pm 0.024)\,M_\odot$.  The mass of
\psr~is at the upper edge of the range observed in mildly recycled
pulsars in double neutron star systems, consistent with the the
recycling hypothesis. It appears that the production of millisecond
pulsars is possible with the accretion of less than $0.2\,M_\odot$.
\end{abstract}


\section{Introduction}\label{sec:intro}

Radio pulsars in binary systems can allow the measurement of neutron
star masses, thereby providing one side of the dense matter equation
of state.  However, to date all pulsars with precisely-measured masses
have been relatively slowly rotating objects (spin period $P$
typically tens of milliseconds) in double neutron star systems or, in
one case, an eccentric pulsar-massive white dwarf system (see Stairs,
2004 for a recent review).\nocite{sta04} These neutron stars all fall
within the narrow mass range of $(1.35 \pm 0.04)\,M_\odot$ established
by \cite{tc99}, with the exception of PSR~B1913+16 at $(1.4408 \pm
0.0003)\,M_\odot$ \citep{wt03} and PSR~J0737$-$3039B at $(1.250 \pm
0.005)\,M_\odot$ \citep{lbk+04}.

Millisecond pulsars ($P < 10$\,ms) are generally thought to have
accreted more matter from a low-mass binary companion and therefore
are expected to be more massive than the slower-rotating recycled
pulsars with massive companions.  However, these low-mass binary
pulsars do not experience measurable relativistic orbital period decay
($\dot{P_b}$) and their typically circular orbits make orbital
precession ($\dot{\omega}$) and time dilation and gravitational
redshift ($\gamma$) difficult to measure; therefore, mass
determinations have been less precise.  There is a statistical
suggestion that millisecond pulsars are indeed more massive than
slower-rotating neutron stars
\citep{ktr94,vbb+01,fck+03}, but with the exception of PSR~J0751+1807
at $(2.1 \pm 0.2)\,M_\odot$ \citep[Nice et al., in
preparation;][]{nss04} and possibly PSR~J1748-2446I in the globular
cluster Terzan~5 \citep{rhs+05}, all are still
consistent with the observed mass distribution of pulsars with massive
companions.

\psr~was discovered during a large-area survey for pulsars at high
galactic latitudes with the 64-m Parkes radio telescope
 \citep{jbvk+03}.  It is a typical low-mass binary pulsar with $P =
 2.95$\,ms, but with several exceptional qualities: its extremely
 narrow pulse profile \citep{ovs+04} and stable rotation allow us to
 measure pulse arrival times with unprecedented precision, and its nearly
 edge-on orbital inclination means the pulsar signal experiences
 strong Shapiro delay.  Measurement of this Shapiro delay gives the
 orbital inclination and companion mass to high precision, and
 when combined with the Keplerian orbital parameters, allows us to measure
 the pulsar's mass.

\section{Observations and Pulse Timing}
\label{sec:obs}

In December 2002, we began observing \psr~with the
Caltech-Parkes-Swinburne Recorder~II \citep[CPSR2, see][]{bai03} at the Parkes radio telescope.  This
instrument samples the voltage signal from the telescope in
each of two 64-MHz wide dual-polarization bands with 2-bit precision.
For these observations, we used the center beam of the Parkes Multibeam
receiver or H-OH receiver and placed the two bands at sky frequencies
of 1341\,MHz and 1405\,MHz.  The raw voltage data were sent to a
dedicated cluster of 30 dual-processor Xeon computers for
immediate analysis.  The two-bit sampled data were corrected for
quantization effects \citep{ja98} and coherently dedispersed into
128 frequency channels in each of 4 Stokes parameters which were then
folded at the topocentric pulse period using the {\sc psrdisp}
software package \citep{van02}.  We recently began observing \psr~with
the Caltech-Green Bank-Swinburne Recorder~II (CGSR2), a clone of CPSR2
installed at the 100-m Green Bank Telescope (GBT).~ As our current GBT data
set covers only a small fraction of the system's orbital phase, we
have not included it in this analysis.

Off-line data reduction and calculation of average pulse times of
arrival (TOAs) were accomplished in the usual manner using the {\sc
psrchive}\footnote{http://astronomy.swin.edu.au/pulsar/software/libraries/}
suite.  Because of roll-off of the anti-aliasing filters, 8\,MHz was
removed from each band edge prior to formation of dedispersed total
intensity profiles, giving a final bandwidth of 48\,MHz per band.  To
avoid averaging over phenomena which vary on orbital timescales,
observations longer than 10 minutes were broken into 10-minute
segments.  Finally, TOAs were calculated by cross-correlation with a
high signal-to-noise template profile, formed by summing a total of 5.4
days of integration in the 1341\,MHz band.  Arrival times with
uncertainty greater than 1\,$\mu$s were excluded from further
analysis.  Our final data set contains 1730 TOAs -- roughly half of
which come from each of our two frequency bands. 

We used the standard pulsar timing package 
{\sc tempo}\footnote{http://pulsar.princeton.edu/tempo/}, along with the Jet
Propulsion Laboratory's DE405 ephemeris, for all timing analysis.  TOAs
were corrected to UTC(NIST).  Using the TOA uncertainties estimated
from the cross-correlation procedure, our best-fit timing model had
reduced $\chi^2 \simeq 1.2$, indicating that our arrival time
measurements are relatively free of systematic errors.  In our final
analysis, these TOA uncertainties were scaled by a factor of 1.1 to
achieve a reduced $\chi^2 \simeq 1$ and improve our estimation of
uncertainties in model parameters.  Because {\sc tempo} estimates
parameter uncertainties based on the assumption that the reduced
$\chi^2$ is unity, and because TOA uncertainties normally must be
scaled by significantly larger factors to satisfy this assumption, it
has become customary to take twice the formal error from {\sc tempo}
as the 1$\,\sigma$ uncertainty to compensate for systematic errors.
We have not followed this practice as our scaling factor is nearly
unity.  Because of the system's low eccentricity ($e$), we used the
ELL1 binary model which replaces the longitude of periastron
($\omega$), time of periastron ($T_0$), and $e$ with the time of
ascending node ($T_{\rm asc}$) and the Laplace-Lagrange parameters $e
\sin \omega$ and $e \cos \omega$ \citep{lcw+01}.  The resulting
parameter values are all consistent with those obtained using the DD
model \citep{dd85, dd86}, though the estimated uncertainties of
several orbital parameters differ significantly.  We give the results
of our timing analysis in Table~\ref{tab:par}.  Although the rms
timing residual could be lowered to 74\,ns by integrating our daily
observations, high time resolution around superior conjunction was
vital in mapping the Shapiro delay of the pulsar.  The weighted rms
residual of only 230\,ns obtained from 10-minute integrations in each
of the two bands is still exceptional.

\subsection{Shapiro Delay and Component Masses}
\label{sec:shapiro}

\begin{figure}
\plotone{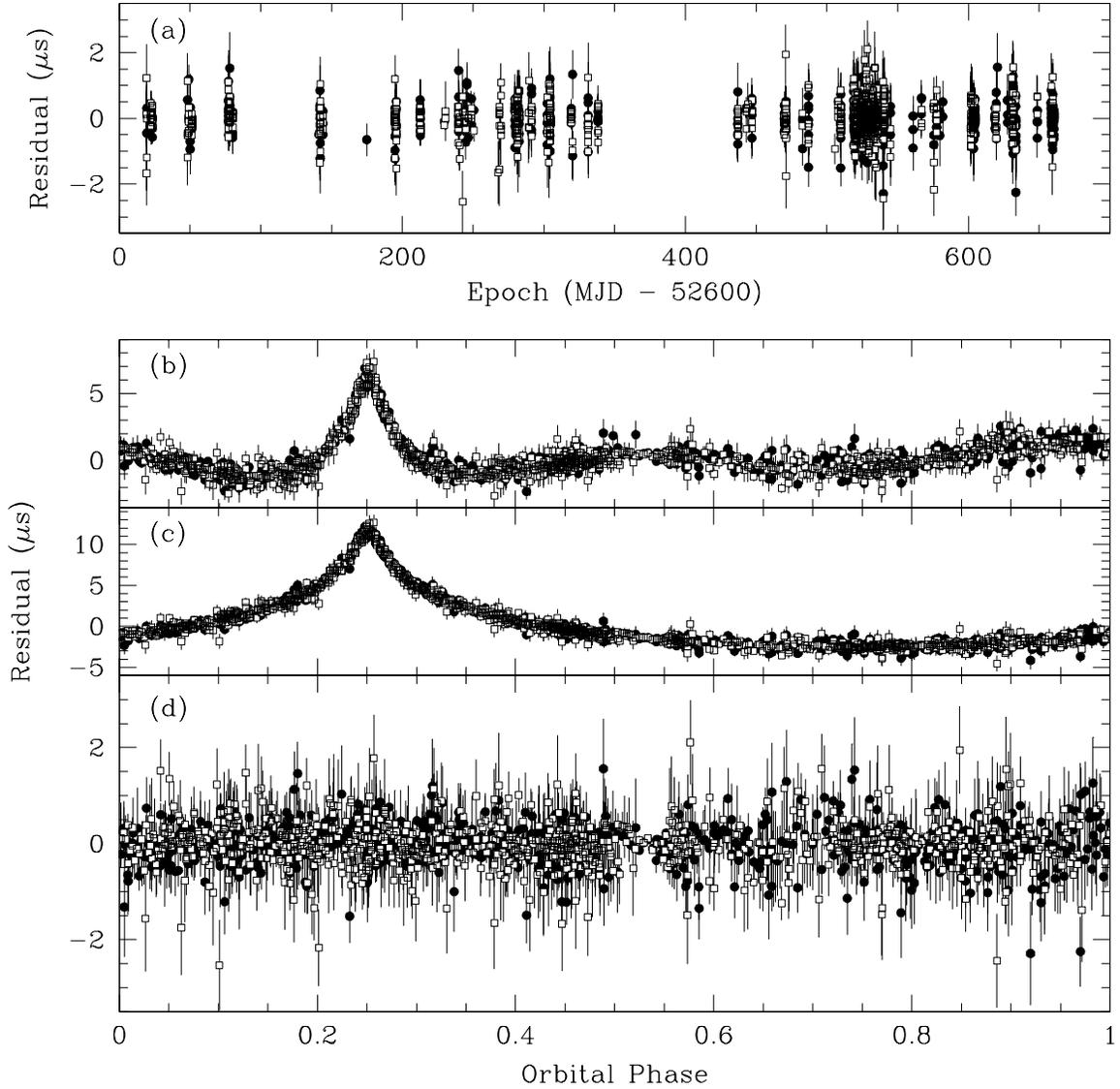}
\caption[High-precision timing residuals for \psr.]{High-precision 
timing residuals for \psr. Filled circles represent TOAs from 
1341\,MHz band, while open squares denote the 1405\,MHz band.  
(a): Residuals vs. observation epoch for
best-fit model taking Shapiro delay fully into account
(Tab. \ref{tab:par}).  (b): Residuals vs. orbital phase for best-fit
Keplerian model.  Some of the Shapiro delay signal is absorbed in an
anomalously large Roemer delay and eccentricity.  (c): Residuals
vs. orbital phase for the best-fit model, but with the companion mass set to zero (i.e. the correct Keplerian orbit, but neglecting Shapiro delay).  (d): Residuals vs. orbital phase for the
best-fit model taking Shapiro delay fully into account
(Tab. \ref{tab:par}).}
\label{fig:1909_residuals}
\end{figure}

\begin{figure}
\plotone{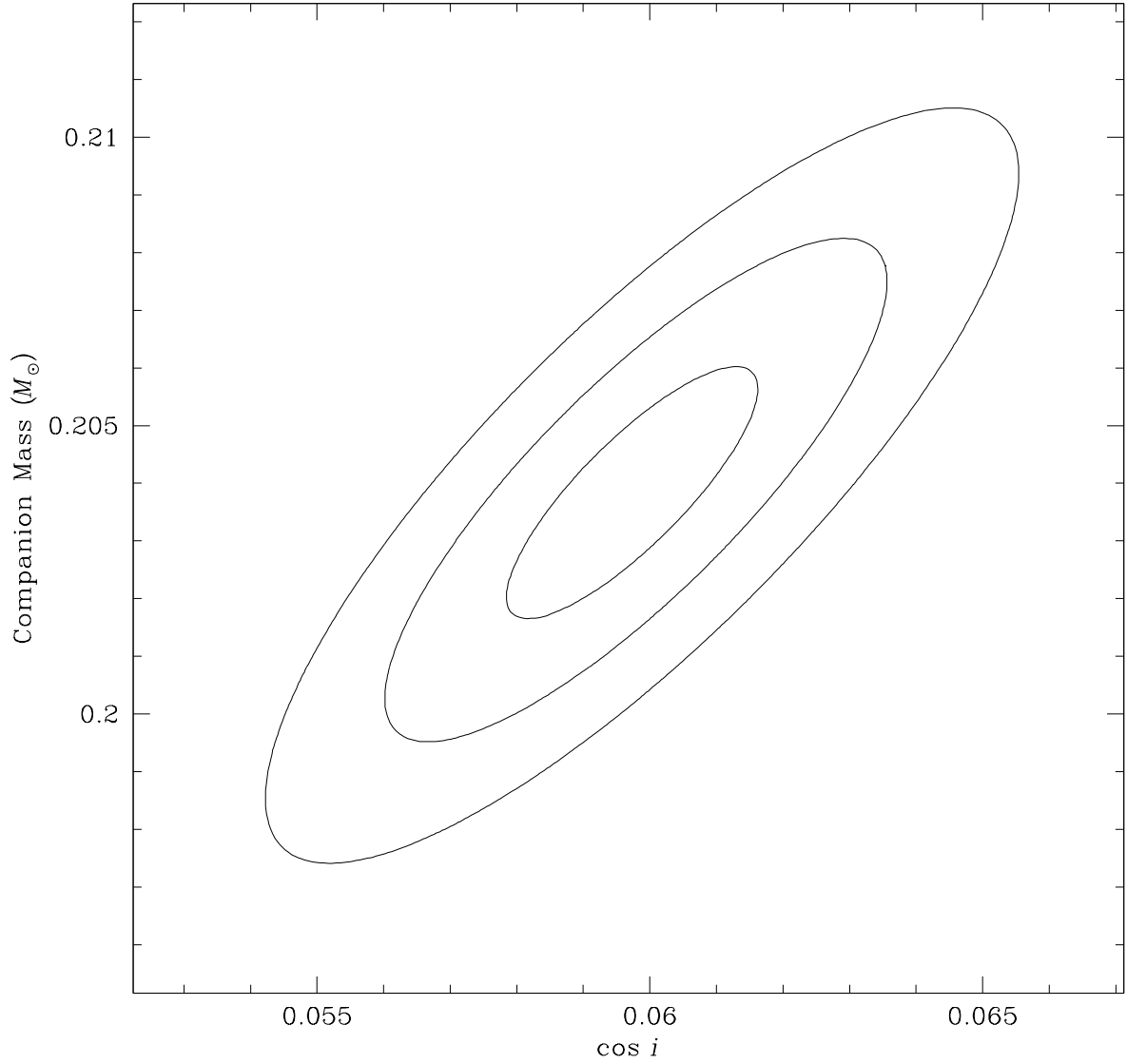}
\caption[Companion mass -- orbital inclination diagram for \psr.]{Companion mass -- orbital inclination diagram for \psr.  Contours show $\Delta \chi ^2 =$ 1, 4, and 9 (1\,$\sigma$, 2\,$\sigma$, and 3\,$\sigma$, or 68.3\%, 95.4\%, and 99.7\% confidence) regions, respectively in companion mass and $\cos i$.} 
\label{fig:1909_m2cosi}
\end{figure}

As shown in Figure ~\ref{fig:1909_residuals}, our timing data display
the unmistakable signature of Shapiro delay.  Measurement of this
relativistic effect has allowed the precise determination of orbital
inclination, $i = (86.58 ^{+0.11}_{-0.10})\arcdeg$, and companion mass, $m_c =
(0.2038 \pm 0.0022)\,M_\odot$.  These values were derived from a $\chi
^2$ map in $m_c$ -- $\cos i$ space (Fig. \ref{fig:1909_m2cosi}), but
are in excellent agreement with the results of {\sc tempo}'s linear
least-squares fit for $m_c$ and $\sin i$.

\begin{figure}
\plotone{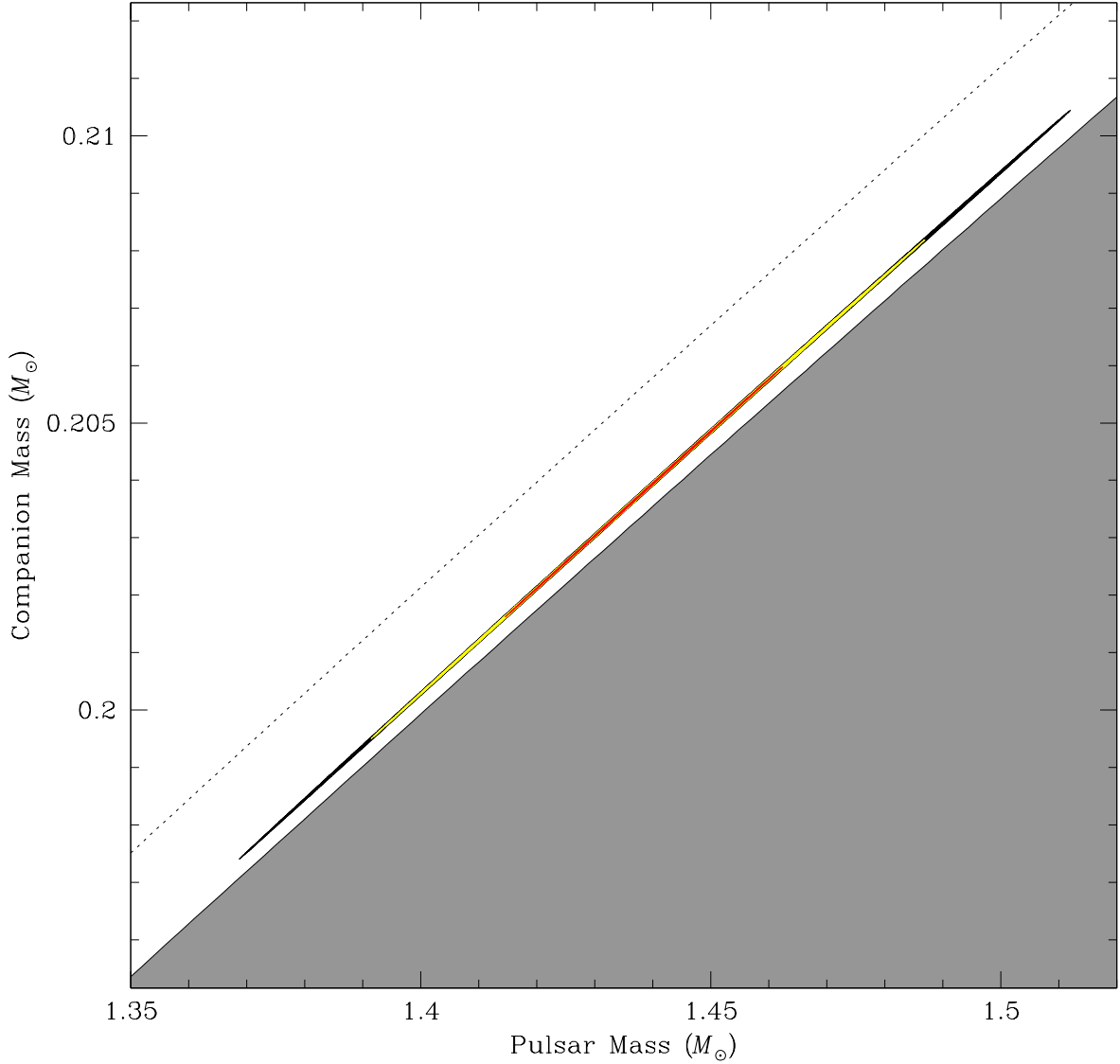}
\caption[Mass -- mass diagram for \psr.]{Mass-mass diagram for \psr.  Red, yellow, and black areas show $\Delta \chi ^2 =$ 1, 4, and 9  (1\,$\sigma$, 2\,$\sigma$, and 3\,$\sigma$, or 68.3\%, 95.4\%, and 99.7\% confidence) regions, respectively in companion mass and pulsar mass.  The grey shaded region is excluded by the mass function and the requirement that $\sin i \le 1$; the dotted line indicates $\sin i = 0.99$ for reference.} 
\label{fig:1909_m1m2}
\end{figure}

Combined with the mass function, our tight constraints on $m_c$
and $i$ determine the pulsar mass, $m_p = (1.438 \pm
0.024)\,M_\odot$ (Fig. \ref{fig:1909_m1m2}).  This result is several
times more precise than the previous best mass measurements of heavily
recycled neutron stars.

We have used Shapiro delay to measure two post-Keplerian parameters
for this system, shape $s \equiv
\sin i = 0.99822 \pm 0.00011$ and range $r \equiv m_c G / c^3 =
(1.004 \pm 0.011)\,\mu{\rm s}$, where $G$ is the gravitational
constant and $c$ is the speed of light.  The value of $\dot{\omega}$
predicted by general relativity is only 0.14 deg yr$^{-1}$, so it will
be many years before a third post-Keplerian parameter can be measured
in this extremely circular system. We note that \psr~has the
smallest eccentricity of any known system, $e = 1.35(12)\times10^{-7}.$

\subsection{Distance and Kinematic Effects}

The measured parallax, $\pi = (0.88\pm0.03)$\,mas gives a distance of
$d_{\pi} = (1.14^{+0.04}_{-0.03})$\,kpc.  We can now calculate the
mean free electron density along the line of sight to
\psr~based on its dispersion measure ($DM$), $\left<n_e\right> \equiv DM/d = DM \pi = (0.0091 \pm 0.0003)\,{\rm cm}^{-3}$.  
The $DM$-derived distance estimate is 0.46\,kpc
\citep{cl02}, indicating that the free electron density along the line
of sight is significantly overestimated by the model.  There are no
pulsars near \psr~on the sky with accurate parallax measurements,
either through pulse timing or interferometry; \psr~will therefore provide an
important constraint on galactic electron density models.

The pulsar's proper motion induces an apparent secular acceleration
equal to $\mu^2 d / c$ \citep{shk70}.  The secular acceleration
corrupts the observed period derivative ($\dot{P}$).  Since the
pulsar's intrinsic spindown rate must be
non-negative, we can obtain an upper distance limit of $d_{\rm max} =
1.4$\,kpc, consistent with the measured parallax.  Conversely, the
measured proper motion and parallax-derived distance allow us to
correct  for the secular acceleration and determine
the intrinsic spindown rate, $\dot{P}_{\rm int}$, which we then use to calculate the characteristic age, $\tau_c = P / (2 \dot{P}_{\rm int})$, and surface dipole magnetic field, $B_{\rm surf} = 3.2 \times 10^{19} (P \dot{P}_{\rm int})^{1/2}\,{\rm G}$ (Tab.\ \ref{tab:par}).  We note that the
acceleration induced by differential Galactic rotation is about two
orders of magnitude smaller than the secular acceleration and
has therefore been neglected.  

Similarly, the secular acceleration induces an apparent $\dot{P_b}$.
Based on the distance and proper motion of \psr, this kinematic
$\dot{P_b}$ is expected to be $\sim 0.5 \times 10^{-12}$ -- about 400
times larger (and of opposite sign) than the predicted intrinsic
value.  We note that if we include $\dot{P_b}$ in our timing model, the
best-fit value is consistent with this prediction but the
significance is low.  Therefore, we have not included $\dot{P_b}$ in
our analysis.  However, in several years, this kinematic $\dot{P_b}$
will give an improved measurement of the pulsar distance \citep{bb96}.

The transverse velocity resulting from the parallax distance and the
measured proper motion is $(200^{+7}_{-6})$\,km\,s$^{-1}$, somewhat
higher than typical for binary millisecond pulsars \citep{tsb+99}.  We
note that, in the absence of the parallax distance measurement, we
would have significantly underestimated the system's velocity.

\section{Conclusions}
\label{sec:conclusions}

We have obtained the most precise mass of a heavily recycled
neutron star through high-precision timing measurements of \psr:
$m_p = (1.438 \pm 0.024)\,M_\odot$.  While \psr~appears to be
more massive than the canonical range for mildly recycled pulsars
($1.35 \pm 0.04\,M_\odot$), its mass is consistent with that of the
original member of this class, namely PSR~B1913+16.  Therefore, it is
unlikely that a clear mass distinction can be made between the
millisecond pulsars with low-mass white dwarf companions and the
mildly recycled pulsars with massive companions.

It is possible that neutron star birth masses are simply not as
homogeneous as the sample of measurements suggested until recently.
The discovery of PSR~J0737$-$3039B in the double pulsar system with
$m_p = (1.250 \pm 0.005)\,M_\odot$ is difficult to reconcile with the
millisecond pulsars PSR~J0751+1807 and PSR~J1748-2446I, with most
probable masses near or exceeding $2\,M_\odot$, based purely on
post-supernova accretion history.  If we assume that the mass of
PSR~J0737$-$3039B (the only well-measured slow pulsar mass and the
lowest measured neutron star mass) is representative of pre-accretion
neutron stars, it appears that the accretion of less than
0.2\,$M_\odot$ is sufficient for spinning pulsars up to millisecond
periods, implying that most of the companion's original mass may be
lost from the system.  On the other hand, the more massive MSPs
suggest that, in some cases, the accreted mass is several times
larger.

Shortly after their discovery, it was proposed that millisecond pulsars
were produced from normal neutron stars that were recycled by the
accretion of matter \citep{acrs82}. However an alternative
hypothesis was put forward by \cite{bg90} in
which millisecond pulsars were created by the accretion induced 
collapse of a white dwarf. In this scenario, the millisecond pulsar
should be less than the Chandrasekhar mass minus the binding energy
that is realized upon collapse of the neutron star. \psr's mass suggests
that the recycling process is more probable.

The timing precision of \psr~with a 64-m class telescope is
extraordinary.  Every 10 minutes, CPSR2 yields two arrival times with
a weighted rms residual of just 230\,ns. By integrating the data
further, we found we could reduce the residual to just 74\,ns, but
there is reason to believe that we can reduce this figure still
further. At present typical integrations are only an hour, and there
is potentially 512\,MHz of bandwidth at 20\,cm available for precision
timing at the Parkes telescope, only 128\,MHz of which is used by the
CPSR2 instrument. Longer observations with four times the bandwidth
could potentially yield arrival times with an integrated rms residual
of just a few tens of ns. Such precision would allow us to probe the
Universe for the signature of supermassive black hole binaries because
of the effect of gravitational waves on timing measurements
\citep{jb03, lbs+03}.

\acknowledgments

We acknowledge S. Anderson, W. van Straten, J. Yamasaki, and
J. Maciejewski for major contributions to the development of CPSR2,
and thank H. Knight for assistance with observations. These data
contain some arrival times from the P456 program of R. Manchester et al.
BAJ and SRK thank NSF and NASA for supporting their research.  The Parkes
telescope is part of the Australia Telescope which is funded by the
Commonwealth of Australia for operation as a National Facility managed
by CSIRO.

\begin{deluxetable}{ll}
\tabletypesize{\scriptsize}
\tablewidth{0pt}
\tablecaption{Improved parameters of the \psr\ system \label{tab:par}}
\tablehead{
\colhead{Parameter} & \colhead{Value\tablenotemark{a}}
}
\startdata
Right ascension, $\alpha_{\rm J2000}$\dotfill &  $19^{\rm h}09^{\rm m}47\fs4379988(6)$  \\
Declination, $\delta_{\rm J2000}$\dotfill  &   $-37\arcdeg44\arcmin14\farcs31841(4)$ \\
Proper motion in $\alpha$, $\mu_{\alpha}$ (mas yr$^{-1}$)\dotfill   &  $-$9.470(11) \\
Proper motion in $\delta$, $\mu_{\delta}$ (mas yr$^{-1}$)\dotfill  & $-$35.76(8) \\
Annual parallax, $\pi$ (mas)\dotfill  &  0.88(3) \\
Pulse period, $P$ (ms)\dotfill & 2.947108021647488(3) \\
Period derivative, $\dot{P}$ (10$^{-20}$)\dotfill & 1.40258(4) \\
Reference epoch (MJD)\dotfill   &    53000.0  \\
Dispersion measure, $DM$ (pc cm$^{-3}$)\dotfill       &       10.39392(6) \\
Binary period, $P_b$ (d)\dotfill     &    1.533449450441(10) \\
Projected semimajor axis, $a \sin i$ (lt-s)\dotfill      &      1.89799117(4) \\
$e \sin \omega$ ($\times 10^{-7}$)\dotfill   &      0.56(18) \\
$e \cos \omega$ ($\times 10^{-7}$)\dotfill    &     $-$1.24(10) \\
Time of ascending node, $T_{\rm asc}$ (MJD)\dotfill  &   53000.4753280898(13) \\
$ \sin i$\dotfill    &      0.99822(11)  \\
Companion mass $m_c$ (M$_{\odot}$)\dotfill    &          0.2038(22) \\
\cutinhead{Derived Parameters}
Pulsar mass $m_p$ (M$_{\odot}$)\dotfill    &          1.438(24) \\
Mass function, $f(m)$\dotfill &  0.0031219531(2) \\
Range of Shapiro delay, $r$ ($\mu$s) \dotfill & 1.004(11) \\
Orbital inclination, $i$ (deg) \dotfill    &  $86.58 ^{+0.11}_{-0.10}$     \\
Orbital eccentricity, $e$ ($\times 10^{-7}$)\dotfill     &    1.35(12) \\
Longitude of periastron, $\omega$ (deg)\dotfill & $ 155.7452858095 \pm 7 $ \\
Time of periastron, $T_0$\dotfill    &   $53001.13873788 \pm 0.03$\\
Parallax distance, $d_{\pi}$ (kpc)\dotfill & $1.14^{+0.04}_{-0.03}$ \\
Transverse velocity, $v_{\perp}$ (km s$^{-1}$)\dotfill & $200^{+7}_{-6}$  \\
Intrinsic period derivative, $\dot{P}_{\rm int}$ (10$^{-20}$)\tablenotemark{b}\dotfill & 0.28(4) \\
Surface magnetic field, $B_{\rm surf}$ $(\times 10^8 \rm{G})$\tablenotemark{b}\dotfill & $0.92^{+0.06}_{-0.07}$ \\
Characteristic age, $\tau_c$ (Gyr)\tablenotemark{b}\dotfill & $16^{+3}_{-2}$ \\
Galactic longitude, $l$ (deg)\dotfill & 359.73 \\
Galactic latitude, $b$ (deg)\dotfill &  $-$19.60 \\
Distance from Galactic plane, $|z|$ (kpc)\dotfill & 0.383(12) \\
\enddata
\tablenotetext{a}{Figures in parenthesis are uncertainties in the last digit quoted.  The formal error calculated by {\sc tempo} is taken as the 1$\,\sigma$ uncertainty, except as described in \S \ref{sec:shapiro}; all quoted uncertainties correspond to the 68.3\% confidence interval.}
\tablenotetext{b}{Corrected for secular acceleration based on measured proper 
motion and parallax}
\end{deluxetable}

\end{document}